\documentclass[12pt,preprint]{aastex}
\usepackage{amsmath}

\def\bea{\begin{eqnarray}}
\def\eea{\end{eqnarray}}
\def\be{\begin{equation}}
\def\ee{\end{equation}}

\newcommand{\showlabel}[1]{} 
\newcommand{\internalcom}[1]{}

\received{}
\accepted{}
\journalid{}{}
\articleid{}{}
\paperid{}
\cpright{AAS}{2011}
\ccc{}

\shorttitle{ $n\ell \rightarrow n\ell'$ Proton-impact cross-sections  }
\shortauthors{Vrinceanu, Onofrio, and Sadeghpour}

\begin{document}

\title{Angular momentum changing transitions in \\ 
proton-Rydberg hydrogen atom collisions}

\author{
D. Vrinceanu\altaffilmark{1}, 
R. Onofrio\altaffilmark{2,3},
and H. R. Sadeghpour\altaffilmark{3}} 

\altaffiltext{1}{Department of Physics, Texas Southern University, Houston, TX 77004, 
USA; daniel.vrinceanu@gmail.com}

\altaffiltext{2}{Dipartimento di Fisica e Astronomia ``Galileo Galilei,'' Universit\`a di Padova, 
Via Marzolo 8, 35131 Padova, Italy, and ITAMP, Harvard-Smithsonian Center for 
Astrophysics, Cambridge, MA 02138, USA; onofrior@gmail.com}

\altaffiltext{3}{ITAMP, Harvard-Smithsonian Center for Astrophysics, 
Cambridge, MA 02138, USA; hrsadeghpour@gmail.com}

\begin{abstract}
Collisions between electrically charged particles and neutral atoms are central for understanding 
the dynamics of neutral gases and plasmas in a variety of physical situations of terrestrial and 
astronomical interest. Specifically, redistribution of angular momentum states within the degenerate 
shell of highly excited Rydberg atoms occurs efficiently in distant collisions with ions. 
This process is crucial in establishing the validity of the local thermal equilibrium assumption 
and may also play a role in determining a precise ionization fraction in primordial recombination. 
We provide an accurate expression for the non-perturbative rate coefficient of collisions between 
protons and H($n\ell)$ ending in a final state H($n\ell')$, with $n$ being the principal quantum 
number and $\ell,\ell'$ the initial and final angular momentum quantum numbers respectively.
The validity of this result is confirmed by results of classical trajectory Monte Carlo simulations.
Previous results, obtained by Pengelly and Seaton only for dipole-allowed transitions $\ell \rightarrow \ell \pm 1$, 
overestimate the $\ell$-changing collisional rate coefficients approximately by a factor of six, and 
the physical origin of this overestimation is discussed. 
\end{abstract}


\section{Introduction}

Collisions between electrons or protons and Rydberg atoms modify the atomic level populations, 
enhancing or suppressing statistical equilibrium, and upsetting selection rules and decay 
channels established by purely radiative transitions. 
Electron-H($n$) collisions are more efficient in transferring energy than p-H($n$) 
collisions, while the latter are far more effective in mixing $\ell$ states. 
Collisional capture of  electrons into highly-excited states is the primary process by
which atoms form in cold and ultracold plasmas, and by which anti-hydrogen atoms are
created in non-neutral magnetized antiproton and positrons plasmas \citep{gabrielse}.
During this stage, high-$\ell$ states are preferentially populated, and subsequent 
collisions with electrons (and ions) populate low-$\ell$ states that have much faster radiative de-excitation rates.

Early theoretical results for proton-hydrogen atom collisions were obtained by \citet{demkov} 
for low quantum numbers, by \citet{percival} using a classical model, and by \citet{abrines} 
employing classical trajectory Monte-Carlo (CTMC) simulations.
Measurements of low $\ell$-mixing rates were performed for beam gas Na$^+$-Na($n \ell$) collisions 
\citep{sun} and theoretical comparison was provided via Floquet analysis \citep{cava}. 
More recently, electron $\ell$-changing collisions were shown to be responsible for 
transferring magneto-optically trapped rubidium Rydberg atoms into high angular momentum states 
\citep{dutta}. In this case, electron $\ell$-mixing efficiency is boosted by the presence of the 
trapping potential allowing for frequent collisions with the Rydberg atoms. 
The $\ell$-changing processes are also important in Zero Electron Kinetic Energy 
(ZEKE) spectroscopy \citep{schlag}.

In a more strict astrophysical setting, collisional physics, both $n$-changing and  $\ell$-changing, 
is critical to understand the validity of the assumptions on which the determination of stellar 
physical parameters - such as effective temperature, density, and chemical abundances - relies  
\citep{mash1,mash2,berg,hill}. The concept of thermodynamical equilibrium applicable in the 
stellar interior - where the mean free paths for photons, electrons, and ions is rather small 
with respect to their average distance - no longer holds near the surface of the star. 
In an intermediate regime, the photons escape with a large mean free path, while electrons and 
ions still maintain small mean free paths for collisions. This is the origin of a peculiar status in 
which, although the photon distribution departs from the equilibrium black-body distribution, the gas 
particles still achieve energy distribution characteristic of thermodynamic equilibrium. 
This allows for the evaluation of  gas particle population via the knowledge of the local temperature 
alone, known as Local Thermodynamic Equilibrium (LTE). 

However, in the upper part of the stellar atmosphere, the chromosphere, the particle densities drop 
so much that even collisions among the gas particles and photons are not enough for maintaining LTE. 
This leads to a breakdown of LTE, with the atomic populations and the ion and electron densities no 
longer determined by the Boltzmann and Saha formulas, and the detailed knowledge of atomic level 
population for each species, including angular momentum states, is instead required \citep{samp}. 
The validity and the breakdown of LTE depend upon the strength of the atomic transitions. 
For transitions having optical depths larger than the optical depth in the surrounding continuum, 
the thickness of the corresponding region is relatively small, allowing one to meaningfully use 
the average values for the entire region. This is not true if the line-forming region is located 
in the chromosphere, where the condition of LTE has to be relaxed.
Collisional excitations are also important for the determination of the primordial helium 
abundance \citep{luri}, the understanding of line formation for early-type stars 
\citep{przy}, and the spectroscopy of planetary nebulae \citep{piph,brok,samu,otsu}. 

Moreover, the redistribution of $\ell$-state populations in H($n\ell)$ in high redshift 
universe can change the spectrum of the primordial recombination at low frequencies, leading to 
emission of successively lower-frequency photons and spectral distortion. The new surveys of the 
cosmic background background, such as the 7-year data integration of the WMAP satellite, and 
the forthcoming data from the Planck Surveyor, promise higher-precision determination of the 
cosmological parameters, in particular the spectral index of scalar perturbations and the 
baryon content of the Universe. These two parameters are directly affected by modifications 
to the cosmological recombination models. Mixing among high-$\ell$ states, for lower redshifts, 
leads to an increase in recombination to the ground state, while $n$-changing collisions suppress 
the emission of recombination epoch photons \citep{chlu1,chlu2}. 

In this work, we provide analytical expressions for $\ell$-mixing rate coefficient 
in collisions between a Rydberg hydrogen atom and a proton, represented by the process:
\[
\mbox{H}(n,\ell) + \mbox{H}^+ \rightarrow \mbox{H}(n,\ell') + \mbox{H}^+.
\]
The theoretical model can treat the more general case of collisions with
an ionic projectile of mass $M$ and charge $Z$. A semiclassical perturbation
theory for $\ell \rightarrow \ell \pm 1$ collisions \citep{peng} is divergent 
for both small and large impact parameters, requiring the introduction of an 
{\it ad-hoc} radial cut-off to regularize the results. 
The model proposed in this paper is non-perturbative and the only assumption made is 
that the projectile moves on a straight line. This is well justified by the fact 
that collisions with large impact parameter have large probability for $\ell$-changing, 
as first shown in \citet{star1}. Moreover, our model predicts that the rates decrease 
roughly as $|\Delta \ell|^{-3}$.

\section{Angular momentum changing transitions}

In our model for collisions between charged particles and Rydberg atoms, we make three basic assumptions.
First, we assume that collisions occur on timescales which are much longer than the orbital motion of the 
Rydberg electron. This implies that the Rydberg electron travels many orbitals during one collision 
in order to appreciably change its angular momentum. Second, we assume that the colliding particle creates 
a weak electric field, yet large enough with respect to the Stark electric field required to lift the 
hydrogenic degeneracy in each Rydberg manifold of principal quantum number $n$. Third, for pure angular 
momentum mixing without energy exchange, it is necessary that collisions occur at large impact parameters. 
Within the above mentioned approximation scheme, the physical picture for $\ell$-changing collisions is the following.
For collisions at large impact parameters, it is reasonable to assume that the charged projectile moves mostly 
undisturbed, along a classical straight line trajectory, creating a weak electric field \citep{star1}. 
Under the influence of this electric field, the Rydberg electron slowly precesses mixing its angular 
momentum state inside the degenerate shell. This collision is so weak that the angular momentum changes 
without any exchange of energy. Because of the long range nature of the Coulomb interaction, the cross 
section for this process can be very large, and indeed we will find that it is logarithmically diverging 
for $\Delta \ell=\pm 1$ transitions, requiring a semiclassical treatment to cure this divergence. 
Based on geometric arguments, it is expected that the cross section scales as
$\sigma \sim \pi n^4 a_0^2$, where $a_0$ is the Bohr radius.
The results introduced in this paper are valid for $n>10$ (semiclassical approximation), for 
projectile velocities smaller than the orbital velocity of the Rydberg electron, which implies an upper 
bound on $n \sqrt{T} < 2.4 \times 10^4 K^{1/2}$, and for densities greater than $N_{\mathrm{crit}}$ as defined by Eq.~(\ref{Ncrit}).

\subsection{Straight-line trajectory time dependent probability}

Within the above mentioned approximations, the time dependent Schr\"odinger's equation can be analytically 
solved for the evolution of the states within a Rydberg shell, subject to the electric potential created by 
the passing charged projectile \citep{star1}. The key step in finding the exact solution is the observation 
that the angular momentum and Runge-Lenz operators generate an SO(4) symmetry group, which in turn
decomposes into the direct product of two rotation groups SO(3)$\otimes$SO(3). 

The cross section for $\ell$-changing collisions within an energy shell with principal quantum number $n$ 
can be written as an integral over the probability of the impact parameter as
\be\label{x-sec}
\sigma^{(n)}_{\ell\rightarrow\ell'} = 2\pi \int_0^\infty P^{(n)}_{\ell\rightarrow\ell'}\;bdb,
\ee
where the probability for making $\ell\rightarrow \ell'$ transitions has the form,
\begin{equation}\label{prob}
P^{(n)}_{\ell\rightarrow\ell'} = (2 \ell' + 1) \sum_{L = |\ell' - \ell|}^{n-1}
(2 L + 1) 
\left\{ \begin{array}{ccc} \ell' & \ell & L \\ j & j & j \end{array}\right\}^2
\frac{(L!)^2 (n-L-1)!}{(n+L)!} ( 2 \sin \chi)^{2L}
\left[C^{(L+1)}_{n-L-1} (\cos \chi)\right]^2
\end{equation}
where $\{ \cdots \}$ is a six-$j$ symbol, $L$ is the vector sum of $\ell$ and $\ell'$, 
$C^{(\alpha)}_n$ is the ultraspherical polynomial, and $j=(n-1)/2$. 
The rotation angle $\chi$ is defined as \citep{star1,star3}
\begin{equation}\label{chi}
\cos \chi = \frac{1 + \alpha^2 \cos(\Delta \Phi \sqrt{1+\alpha^2})}{1+\alpha^2}
\end{equation}
where $\Delta\Phi$ is the azimuthal angle swept by the projectile, such that in a complete
collision $\Delta\Phi=\pi$, and $\alpha$ is the scattering parameter
\begin{equation}\label{alpha}
\alpha = \frac 32 Z \frac{n \hbar}{m_e v b}
\end{equation}
where $m_e$ is the electron mass and $v$ the initial velocity of the colliding projectile. 
The scattering parameter is directly related to the ratio between the maximum angular momentum 
allowed by the Rydberg electron and the initial angular momentum of the charged projectile.
A complete derivation of the above expressions can be found in \citet{star1,star3}. 

The probability (\ref{prob}) depends on the impact parameter $b$ and the projectile velocity $v$ 
through $\alpha$, such that the cross section in Equation (\ref{x-sec}) can be rewritten as
\[
\sigma^{(n)}_{\ell\rightarrow\ell'} = \frac{9\pi}2 \left(\frac{Z n \hbar}{m_e v}\right)^2 I^{(n)}_{\ell\rightarrow\ell'}\;,
\]
where the velocity independent integral factor
$I^{(n)}_{\ell\rightarrow\ell'}$ is determined by the initial and final states as:
\begin{equation}\label{integral}
I^{(n)}_{\ell\rightarrow\ell'} = \int_0^\infty P^{(n)}_{\ell\rightarrow\ell'}(\alpha)\; \frac{d\alpha}{\alpha^3}
\end{equation}

Since the collision probability scales as $P^{(n)}_{\ell\rightarrow\ell'} \propto \alpha^{2|\ell - \ell'|}$, 
and the rotation angle $\chi \sim 2\alpha$ for small $\alpha$,  the integral factor $I^{(n)}_{\ell\rightarrow\ell'}$,
and therefore the cross section,  diverges for collisions in which $|\Delta \ell|=1$, i. e. the 
{\sl dipole allowed} transitions. The logarithmic singularity in the cross section for large 
impact parameters is well known \citep{peng} and is a reflection of the impact parameter and 
Born approximation. This difficulty has been addressed in \citet{peng} by considering either 
many-body collective effects, such as the plasma screening effects or Debye shielding, or the 
Rydberg natural linewidth, that limits the duration of $\ell$-mixing collisions.

\subsection{Semiclassical probability}

The non-perturbative probability  $P^{(n)}_{\ell\rightarrow\ell'}$ in (\ref{prob}) can be computed with 
the desired numerical precision for a range of quantum numbers, typically $n\leq 60$, before becoming
unstable due to the oscillatory behavior of the ultraspherical polynomials. 
In order to carry out calculations efficiently and analytically beyond intermediate Rydberg principal 
quantum numbers, we use the asymptotic limiting forms of the six-j coefficient and 
the ultraspherical polynomials, $C^{(\alpha)}_n$, and obtain analytical expressions for 
the rate coefficients in the limit of large quantum numbers. A derivation of these results is
detailed in the Appendix.
The probability for $\ell$-mixing transitions in this semiclassical (SC) limit can be written as
\begin{equation}
\label{probSC}
P^{SC}(\ell/n, \ell'/n, \chi ) =
\frac{2 \ell'}{\pi \hbar n^2 \sin\chi}
\left\{
\begin{array}{ccl}
0 & \mbox{,  if } & |\sin \chi | < |\sin (\eta_1 - \eta_2)|\\
\vrule height30pt depth10pt width0pt {\displaystyle K(B/A) \over \displaystyle \sqrt{A}} & \mbox{,  if } &  |\sin \chi |  > |\sin (\eta_1 + \eta_2)|\\
\vrule height30pt depth10pt width0pt {\displaystyle K(A/B) \over \displaystyle \sqrt{B}} & \mbox{,  if } &  |\sin \chi |  < |\sin (\eta_1 + \eta_2)|\\
\end{array}\right.
\end{equation}
where $K$ is the complete elliptic integral, $A = \sin^2\chi - \sin^2(\eta_1 - \eta_2)$,
$B = \sin^2(\eta_1 + \eta_2) - \sin^2(\eta_1 - \eta_2)$, $\cos \eta_1 = \ell/n$, and $\cos \eta_2 = \ell'/n$.
The same result was obtained by \citet{star}, based on calculating the 
overlap between volumes in the classical phase space.

Notice that the scattering probability is zero for $\sin \chi < |\sin(\eta_1 - \eta_2)|$, 
which implies a sharp cutoff for $\chi$, and consequently for the minimum allowed value of $\alpha$ 
for which the transition is possible, $\alpha_{\rm min}$.  
The integral \[I^{\mbox{\tiny SC}}(\lambda, \lambda') = \int_{\alpha_{\rm min}}^\infty 
P^{SC}(\lambda, \lambda', \chi )\;\frac{d\alpha}{\alpha^3}\]
can be obtained in the semiclassical approximation and is divergence-free even for 
$|\Delta \ell|=1$ transitions. Moreover, the calculation of the integral simplifies for 
small angular momentum transfers $\Delta \ell$. In this case, Equation (\ref{expansion}) is 
obtained by writing $\cos \eta_1 = \ell/n$ and $\cos\eta_2 = (\ell + \Delta\ell)/n$ and 
calculating the Laurent series expansion in powers of $\Delta\ell$. 
In this process, we take advantage of the fact that the elliptic function at origin is $K(0) = \pi/2$
and that $\sin\chi  \approx 2 \alpha$ for small $\alpha$.
This simplified expression for transition probability can be easily integrated to
obtain a non-perturbative analytical expression for the corresponding cross sections 
and rate coefficients, for any $\Delta \ell$.

\begin{figure}
\plotone{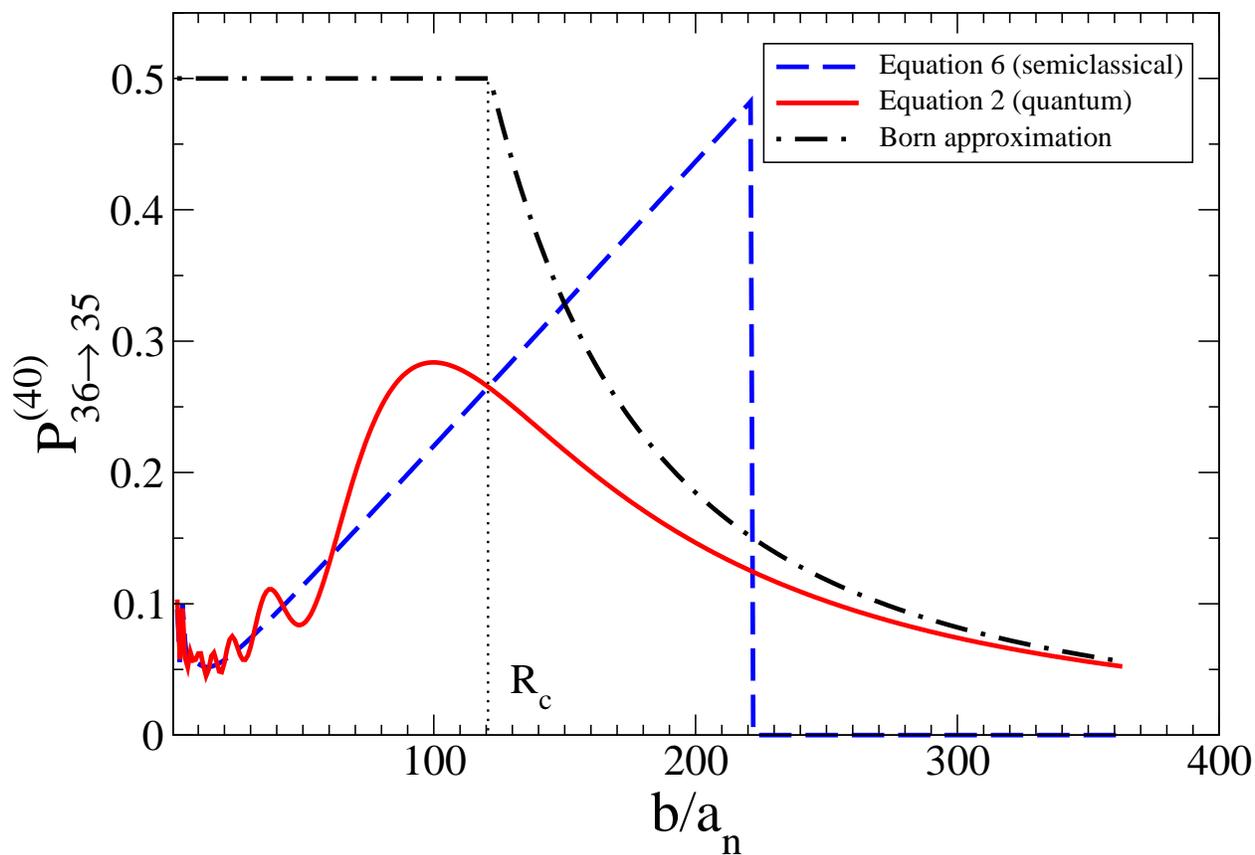}
\caption{Probabilities for a dipole allowed transition in the $n=40$ degenerate hydrogen manifold, 
for a $36 \rightarrow 35$ transition, as a function of the scaled impact parameter $b/a_n$, for 
a fixed projectile velocity $v=0.1$ in atomic units, $a_n=n^2 a_0$. Exact quantum probability (red), 
semiclassical (blue), and Born approximations (black) are compared. The dotted line marks the position 
of the inner cut-off radius used in the Born approximation in \citet{peng}.}
\label{fig1}
\end{figure}

In Figure \ref{fig1}, the collision probability $P^{(n)}_{\ell\rightarrow\ell'} 
\propto \alpha^{2|\ell - \ell'|}$  is plotted versus the impact parameter $b$, scaled 
in units of the Rydberg size $a_n=n^2a_0$, for 
different approximation schemes and a specific dipole ($\Delta \ell=1$) transition 
($n,\ell=40,36\rightarrow n,\ell'=40,35$). 
We can compare the Born approximation used in \citet{peng}, the expression in (\ref{prob}), and 
its semiclassical approximation given by Eq. \ref{probSC}. The result from a direct calculation of 
Equation (\ref{prob}) contains oscillations around the base line formed by semiclassical approximation 
Equation (\ref{probSC}), for most of the impact parameter range. Figure \ref{fig1} shows that the 
probability for angular momentum mixing grows linearly with the impact parameter within the semiclassical 
approximation. This is a counterintuitive conclusion, since one would expect that the ability of a 
projectile to induce changes in a target atom would decrease as the impact parameter increases. 
Due to long range forces and zero energy exchange, angular momentum changing collision are most 
effective at large impact parameters, even for separation of several hundreds Rydberg atom radii. 
The semiclassical approximation falls abruptly to zero at a critical impact parameter given by 
the condition $|\sin\chi|<\sin(\eta_1 - \eta_2)$ in Equation (\ref{probSC}). The quantum expression 
in Equation (\ref{prob}) has a gradual decrease to zero, eventually merging with the Born result 
for sufficiently large $b$, as $b^{-2}$, or $\alpha^2$. 
This is the source of the logarithmic singularity discussed for Equation (\ref{integral}).

The transition probability in \citet{peng} is inversely proportional to $b^2$, posing two 
difficulties: at small $b$ the probability diverges and therefore it cannot be properly 
normalized, and at large $b$ it leads to a divergent cross section. 
Therefore two cut-off parameters are required. A short cut-off $R_c$ is introduced at the point 
where the transition probability is 0.5, under the assumption that for $b < R_c$ the probability 
has fast oscillations that average to 0.5, replacing it by a constant. 
The more serious divergence at large impact parameters is cured by the
second cut-off, that takes into account either the finite density of the medium (plasma) where 
such collisions occur or the finite time of these collision due to the radiative lifetime of 
the Rydberg atom. Figure \ref{fig1} shows that the Born cross sections overestimate the quantal 
and semiclassical cross sections. Our quantum calculation leads to divergent cross sections only 
for $\Delta\ell = \pm1$, while the semiclassical approximation provides finite results for any 
transitions, without the need for a cut-off parameter.

Further simplifications, accurate for most cases of interest, occur if the scaled angular momentum 
$\ell/n$ in the semiclassical formula (\ref{probSC}) is assumed to be a continuous parameter with values 
ranging between 0 and 1, and the the probability is expanded in powers of $|\Delta \ell|/n$.
In this case the integral factor can be calculated term by term, starting with $1/|\Delta \ell/n|^3$, to get
\begin{equation}\label{expansion}
\begin{split}
I^{(n)}_{\ell\rightarrow\ell'} &\approx \frac 1n I^{\mbox{\tiny SC}}(\ell/n, \ell'/n) = \\
&\frac 1n \frac{\ell_<}{\ell} \left\{ 
\frac{2 \left[ 1 - (\ell_</n)^2\right]/3}{|\ell'/n - \ell/n|^3} + 
\frac{\left[ 1 - 3 (\ell_</n)^2\right]/(3 \ell_</n)}{|\ell'/n - \ell/n|^2} + 
{\cal O}\left( \frac 1{|\ell'/n - \ell/n|}\right)
\right\}
\end{split}\end{equation}
where $\ell_<= \mathrm{min}(\ell,\ell')$.

\begin{figure}
\plotone{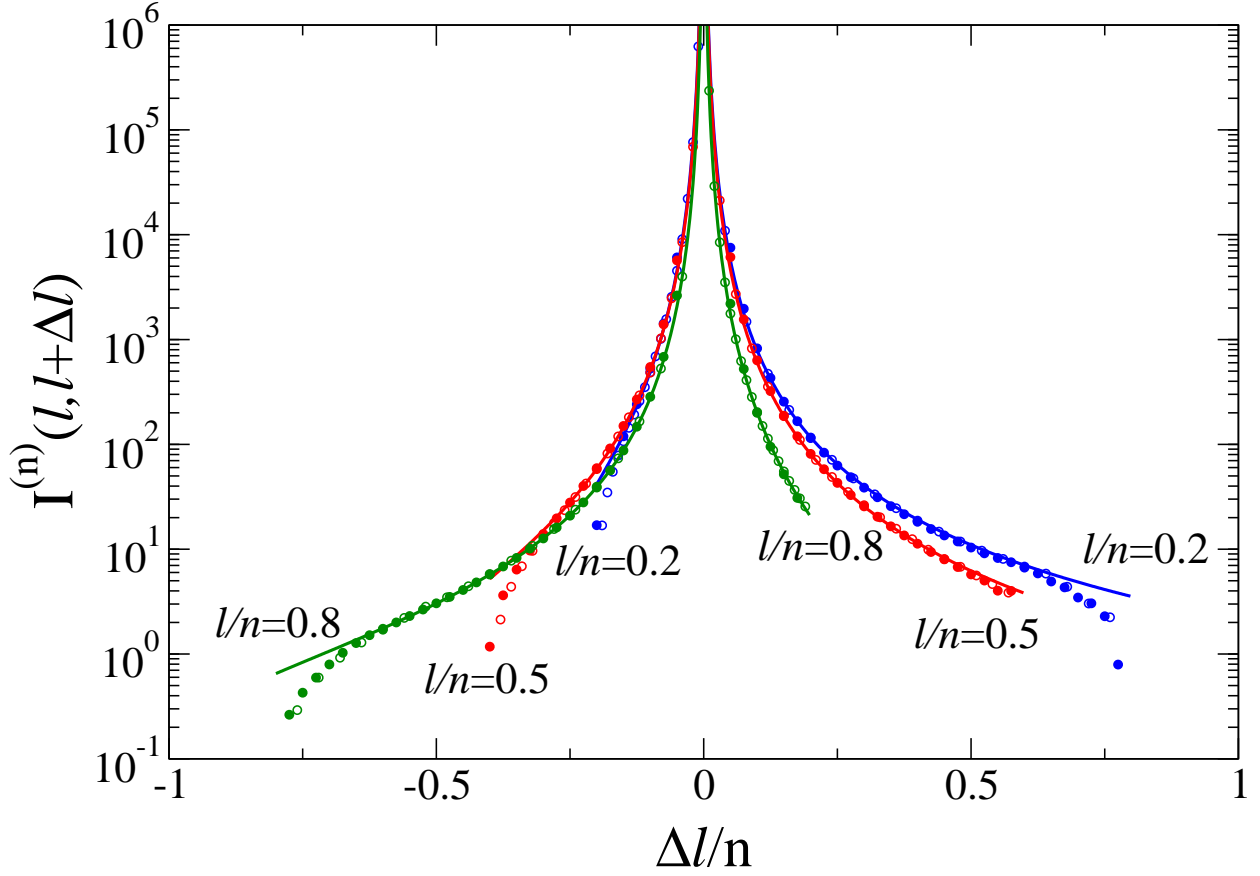}
\caption{Integral factor $I^{(n)}_{\ell\rightarrow\ell'}$ in Equation (5) calculated rigorously and 
using a simplified semiclassical expression, as a function of the scaled angular momentum 
transfer. The continuous line, the empty circles, and the full circles are obtained from 
the expressions in Equations (7),(6), and (2), respectively.}
\label{fig2}
\end{figure}

In Figure \ref{fig2}, we compare the integral factor for transitions inside the $n=40$ shell, starting 
from $\ell =$ 8, 20, and 32 and going to all possible final angular momenta. Three methods are used in 
this calculation: integration of the quantum formula (\ref{prob}), direct integration of the semiclassical 
result (\ref{probSC}), and the asymptotic expansion (\ref{expansion}), dominating small angular momentum 
transfers. The simple expression (\ref{expansion}) is quite effective even for larger transfers. 
The next term in the expansion, proportional to $1/\Delta\ell$, is easily obtained, but it does not 
significantly improve the agreement with quantum calculations.

\subsection{Rate Coefficients}\label{rates}

The rate coefficient for $\ell$-changing transitions in contact with ions having a Maxwell-Boltzmann
distribution $f_{\mbox{\scriptsize MB}}$, is
\[
q_{n\ell\rightarrow n\ell'} = \int v \;\sigma^{(n)}_{\ell\rightarrow\ell'} \;f_{\mbox{\scriptsize MB}}(v) dv = 
\left(\frac{3 Z n \hbar}{m_e}\right)^2 \sqrt{\frac{\pi M}{2 k_B T}} \; I^{(n)}_{\ell\rightarrow\ell'}
\]
where $M$ is the reduced mass of the ion-atom system, and $T$ the temperature of the ions, hereafter assumed to 
be expressed in Kelvin.

By using the simplified semiclassical expression (\ref{expansion}), the collisional rate coefficient for
ion induced $\ell\rightarrow\ell'$ transitions is 
\begin{equation}\label{VOS}
q_{n\ell\rightarrow n\ell'}(T) = 
3\sqrt{\frac{\pi}{2}}\; 
\frac{\hbar^3}{m_e^2 e^2}\;
\sqrt{\frac{m_e e^4}{\hbar^2 k_B T}}\;
Z^2 \sqrt{\frac{M}{m_e}}\;
\frac{n^2 \left[ n^2(\ell + \ell') - \ell_<^2 (\ell + \ell' + 2|\Delta \ell|) \right]}
{(\ell+1/2) |\Delta \ell|^3}.
\end{equation}
The rate coefficient in $\mbox{cm}^3/\mbox{s}$  is then
\begin{equation}\label{rate}
q_{n\ell\rightarrow n\ell'} = 1.294\times 10^{-5} \sqrt{\frac{M}{m_e}} \frac{Z^2}{\sqrt{T}}\;
\frac{n^2 \left[ n^2(\ell + \ell') - \ell_<^2 (\ell + \ell' + 2|\Delta \ell|) \right]}
{(\ell+1/2) |\Delta \ell|^3}
\end{equation}
As expected, this rate scales as $n^4$ with the principal quantum number and rapidly drops with the increase
in angular momentum change, as $1/|\Delta \ell|^3$, similarly to the $\Delta \ell >1$ fitting formula in 
Equation (4.112) of \citet{beig}. The maximum rates are obtained for the dipole allowed transitions, 
$\ell \rightarrow \ell \pm 1$. 
 
It is instructive to compare our two-body rate coefficient with the results of 
\citet{peng}, which only accounts for dipole-allowed transitions,
\be\label{PS}
\begin{split}
q^{PS}_{n\ell}(T) & = 
\sqrt{\frac{\pi}{2}}\; 
\frac{\hbar^3}{m_e^2 e^2}\;
\sqrt{\frac{m_e e^4}{\hbar^2 k_B T}}\;
Z^2 \sqrt{\frac{M}{m_e}}\; D_{n\ell}^{PS}  \times\\
& \ln(10)
\left( \frac{1-\gamma}{\ln(10)} + \log \frac{k_B^2 m_e }{2\pi e^2 \hbar^2} +
\log\frac{T m_e}{D_{n\ell}^{PS} M} + \log \frac{T}{N_e}\right)
\end{split}
\ee
where $\gamma = 0.577216$ is the Euler constant, the coefficient is 
$D_{n\ell}^{PS} = 6 n^2 (n^2-\ell^2-\ell-1)$, and $N_e$ is the electron density expressed in cm$^{-3}$.

Our rate coefficient in Equation (\ref{VOS}), summed over the two dipole transitions is
\be
q_{n\ell}(T) = 2
\sqrt{\frac{\pi}{2}}\; 
\frac{\hbar^3}{m_e^2 e^2}\;
\sqrt{\frac{m_e e^4}{\hbar^2 k_B T}}\;
Z^2 \sqrt{\frac{M}{m_e}}\; D_{n\ell}
\ee
where the coefficient in this case is $D_{n\ell} = 6 n^2 (n^2-\ell^2 -1/4\ell)$. 
The ratio $q_{n\ell}(T)/q^{PS}_{n\ell}(T)$ becomes a slowly varying function of $n$ and $\ell$.
For example, for $(T=3,000$ K, $N_e = 300$ cm$^{-3})$ this ratio is 6.98 for 
($n=50, \ell=48$), and 6.20 for ($n=100, \ell=98$). 
The reason for which the Pengelly and Seaton formula overestimates the angular 
momentum changing rates lies in the Born approximation.

In order to compare the $\ell$-changing collisions with radiative processes, the quantity 
$q_{n\ell} \tau_{n\ell}$ was introduced in \citet{peng}, where $q_{n\ell} = 
q_{n\ell\rightarrow n\ell+1} + q_{n\ell\rightarrow n\ell-1}$  and $\tau_{n\ell}$ is 
the radiative lifetime, which is approximately given by $\tau_{n\ell} \approx 10^{-10} n^3 \ell^2$ s. 
This quantity is the inverse of a critical density above which $\ell$-changing collisions 
are faster than radiative decay from level $n\ell$, which in our case is expressed through the 
relationship (in $\mbox{cm}^{-3}$)  

\begin{equation}\label{Ncrit}
N_{\mathrm{crit}} = (q_{n\ell} \tau_{n\ell})^{-1} = 
73.7 \;\frac{\sqrt{T/10^4 K}}{Z^2}\sqrt{\frac{m_e}{M}}\frac{(n/40)^{-9}}{\left(\ell^2/n^2\right)\left(1-\ell^2/n^2\right)}
\end{equation}


If scaled down by a factor of approximately six, Equation (\ref{Ncrit}) agrees with previous Born results in \citet{peng}.

We may also compare the rates for angular momentum mixing with other processes in plasmas of 
astrophysical relevance, as discussed in \citet{dalgarno}.
Firstly, the angular momentum changing rate in collisions with electrons is smaller 
than the rate in collision with protons by factor of $(m_p/m_e)^{1/2}\simeq 43$.
The energy changing rate due to proton collision is smaller than the 
angular momentum changing rate by $(\Delta n/n)^3$. This factor comes 
from the assumption of straight line trajectory, which implies a transition probability 
proportional to the square of the dipole matrix element. The matrix elements for in-shell 
transitions are greater than those for inter-shell transition by approximately the same factor.

\begin{table}
\begin{center}
\caption{\label{table1}
Rates in $s^{-1}$ for processes occurring in a plasma with number density $10^4$ cm$^{-3}$ and
various temperatures}
\begin{tabular}{crrr}
\tableline
\tableline
Process & 10,000 K & 1,000 K & 25 K \\
\tableline
H$^+$ angular momentum changing           & 6362 & 20119 & 127241 \\
Electron de-excitation                    & 5725 & 840   & 38.9   \\
Electron excitation                       & 14.7 & 21.6  & 40     \\
Radiative relaxation                      & 2442 & 2442  & 2442   \\
\tableline
\end{tabular}
\end{center}
\end{table}

The electron excitation rate coefficient depends as $\sim T^{5/6} n^{14/3}$ on the temperature $T$
and $n$, while the de-excitation rate coefficient depends as 
$\sim T^{-1/6} n^{8/3}$ \citep{pohl}. Therefore for low enough temperatures, these processes will be 
dominated by the proton induced $\ell$-mixing. The radiative relaxation rate,
represented by the inverse of the radiative life time $\tau_{n\ell} \approx 10^{-10} n^3\ell^2$s,
does not depend on temperature or density. A critical electron (or proton) density can be 
defined such that the proton induced $\ell$-mixing rate is faster than the radiative rate \citep{peng}. 
A near-resonant charge transfer is most efficient at projectile velocities matching the velocity of Rydberg 
electrons and at impact parameters in the range of $(5-10) n^2$ \citep{smit}. 
This process is therefore unlikely for $\ell$-changing collisions which occur with large probability 
at non-velocity matching conditions and much larger impact parameters, see Fig. 1. 
Our CTMC simulations confirm that, in the range of parameters considered here, the number of trajectories 
resulting in charge transfer are a small fraction (less than $10 \%$) of the total number of $\ell$-changing trajectories.
We compare rates for these processes at different temperatures in Table \ref{table1}, showing that 
the proton induced $\ell$-changing is the fastest process over a wide range of parameters.

\section{Monte Carlo Simulations}

In order to confirm the range of validity of our semiclassical results in Sec. \ref{rates}, we perform 
numerical simulations of proton (projectile) and H($n\ell$) (target) collisional $\ell$-mixing and calculate 
the rate coefficients for these collisions with the CTMC techniques. The earliest application of CTMC technique 
to proton - hydrogen atom collision is reported in \citet{abrines}, who calculated the classical charge transfer 
and ionization cross sections in $p$ - H$(n=1)$ collision, with the initial ground state kinetic energy sampled 
from a microcanonical distribution. We similarly use a microcanonical population for the Rydberg atoms with 
generic principal quantum number $n$ and angular momentum quantum number $\ell$. The velocities of the projectile 
ion and the center-of-mass of the Rydberg atoms are sampled from a Boltzmann-Maxwell distribution at temperature $T$. 
The collision is started at $t= -t_{max}$ and ends at $t=t_{max}$ where $t_{max}= \eta b/v_{12}$, where 
$\eta =4$ is chosen in our simulations, and $v_{12}$ is the relative velocity between the projectile and Rydberg atom. 
We perform the simulations at temperatures large enough such that the propagation time is manageable, but 
small enough such that energetic collisions could be neglected. All 18 conjugate degrees of freedom 
are propagated in time with a time-adaptive implicit Runge-Kutta integration method of forth order. 
The selection of trajectories for the calculation of rate coefficients is achieved 
by ensuring conservation of the total energy and by maintaining the on-shell energy,  
{\it i. e.} $|n'-n| < 0.5$ where $n'$ is the effective quantum number $n' = 1/\sqrt{-2 E'}$ 
for both the target, or the charge transferred atom, at the end of the collision. 
If the angular momentum of the direct or charge transferred Rydberg atom at the end
of the collision is in the interval $(\ell', \ell'+1)$, then the trajectory is
counted as a collision with final angular momentum $\ell'$.

Care must be exercised in the choice of impact parameter distribution, considering the large range
of $b$ values which are required for effective angular momentum mixing, as illustrated in Figure \ref{fig1}.
According to the model, the  probability for a change of $\Delta\ell$ grows linearly with $b$ up to 
a maximum impact parameter $b_{max} = 3n^2\epsilon/(v|\Delta\ell|)$, and is zero for $b > b_{max}$. 
There are two options to address this problem; one is using the importance sampling, 
{\it i. e.} to try to get a distribution that resembles the integrand, using more points 
where the integrand is large, or the option of using stratified sampling, in which 
the integration domain is partitioned into non-equal slices, each having uniform probability distribution. 
For the stratified sampling at a given $v_{12}$, the $b$ space is divided into intervals 
$D_1$, $D_2, \dots, D_k$, where each segment  is defined by 
${3n^2\epsilon}/{(k + 1/2)} \le b v_{12} \le {3n^2\epsilon}/{(k - 1/2)}$.
If our model is correct, a collision with impact parameter in a segment $D_k$ can lead only
to collisions for which $|\Delta\ell| \le k$.  

\begin{figure}
\plotone{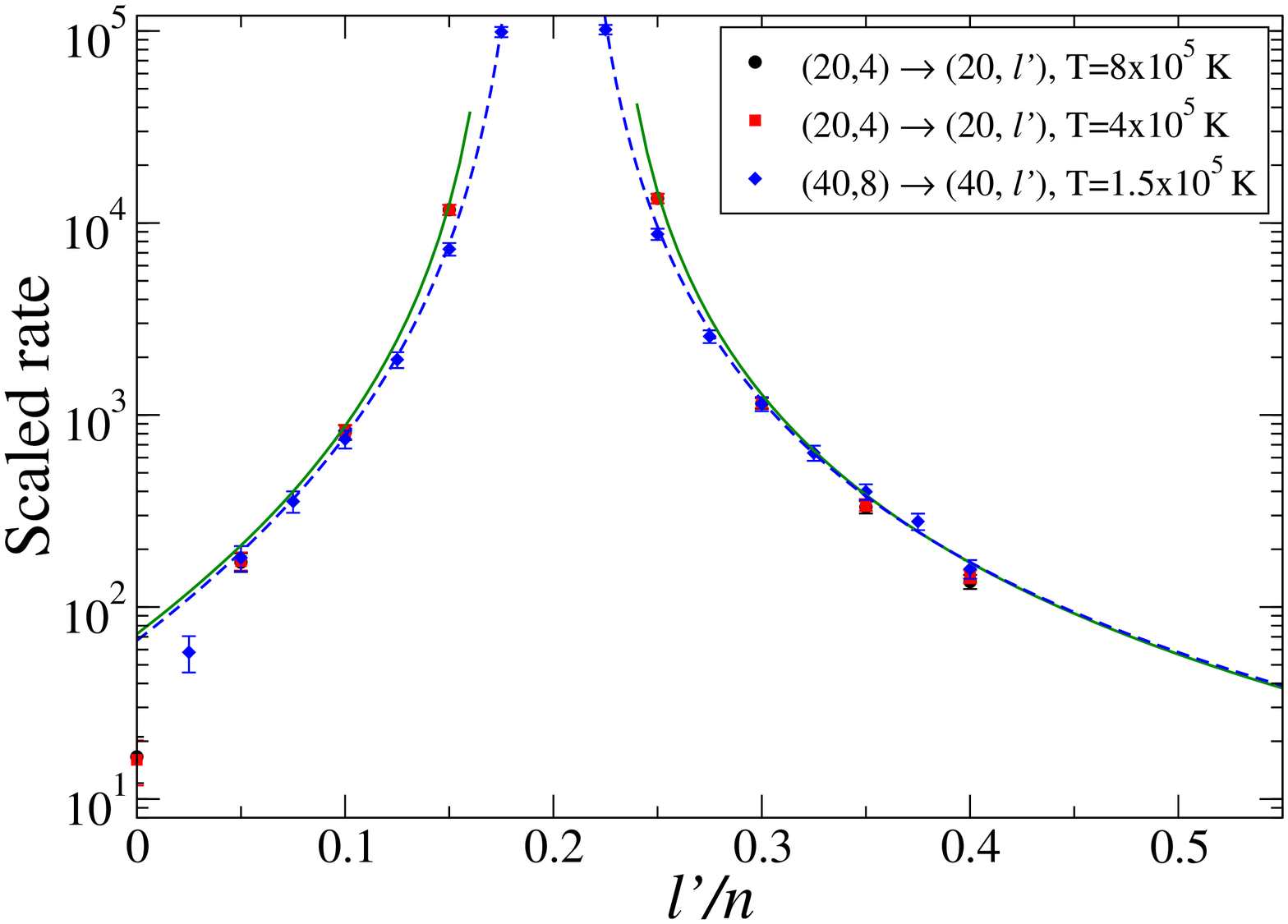}
\caption{CTMC calculation of angular momentum mixing rates scaled by $6 n \sqrt{\pi M/(2 k_B T)}$,
as a function of the scaled final angular momentum. Dots, squares, and diamonds are the results 
for the three cases discussed in the text, with their statistical errors. Solid lines are predictions 
given by Equation (\ref{calR}), integrated over the same final angular momentum bins used in simulations.}  
\label{fig3}
\end{figure}

We have performed simulations for 3 cases: 
(a) $n=20$, $\ell=4$, $T=800,000$ K, 
(b) $n=20$, $\ell=4$, $T=400,000$ K, and 
(c) $n=40$, $\ell=8$, $T=150,000$ K.
For cases (a) and (b), we run simulations for segments $D_1$ to $D_{10}$, and $D_1$ to $D_{19}$ for 
case (c), with 40,000 trajectories within each segment. Figure \ref{fig3} shows the results of the simulations.

In order to represent the results from these cases on the same graph, we scale the rates by 
$6 n \sqrt{\pi M/(2 k_B T)}$, and the angular momenta scaled by $n$, $\lambda = \ell/n$, 
$\lambda' = \ell'/n$. By scaling with the same factor above the rate coefficient, Equation (\ref{rate}) can be 
written, in atomic units, as 
\begin{equation}\label{calR}
{\cal R}(\lambda, \lambda') = \frac{3n}{2} I^{(n)}_{\ell\rightarrow\ell'}=
\frac{\lambda_{<}}{\lambda}\left[\frac{1 - \lambda_{<}^2}{|\Delta \lambda|^3} 
+ \frac{1 - 3 \lambda_{<}^2}{2 \lambda_{<} |\Delta\lambda|^2}\right]\;
\end{equation}
which does not depend on temperature, and depends on $n$, $\ell$ and $\ell'$ only through $\lambda$ and $\lambda'$.

Figure \ref{fig3} shows that the $\ell$-changing rate is divergent at $\ell'/n = 0.2$, because this
corresponds to $|\Delta\ell|\rightarrow 0$, and decreases as $1/|\Delta\ell|^3$ for larger $\Delta\ell$.
The agreement between the simulated cases and the prediction of the simple formula (\ref{rate})
confirms the validity of the semiclassical model. The two curves are slightly different at
small $\Delta\ell$ because they actually represent the scaled rate (\ref{calR}) integrated over
the finite bins used in the simulations. The value of the scaled bin $\delta\ell'/n$ is different 
for case (c). We have also verified through CTMC simulations that $\ell$-changing rates for 
electron-Rydberg H($n\ell)$ are smaller with respect to the proton ones by a ratio 
$\sqrt{m_p/m_e} \simeq 40$, as expected from Equation (\ref{VOS}).

\section{Conclusions}

Analytical expressions for the rate coefficient of $\ell$-changing collisions between protons and H($n\ell)$ 
Rydberg atoms have been derived using a non-perturbative approach free of divergences, plaguing previous 
results for the dipole-allowed $\ell$-mixing collisions. The results have been compared to CTMC simulations 
over a range of temperatures of astrophysical interest. The dipole-allowed $\ell$-changing collision 
coefficient evaluated in \citet{peng} overestimates the corresponding rates by about an order of magnitude.

\acknowledgments
DV is grateful to Texas Southern University High Performance Computing Center for 
making the necessary computational resources available.  This work was partially 
supported by the National Science Foundation through a grant for the Institute for 
Theoretical Atomic, Molecular Physics at Harvard University and the Smithsonian 
Astrophysical Laboratory.

\section*{Appendix: Derivation of Equation (6)}

In this section, the semiclassical expression (\ref{probSC}) is derived from Equation (\ref{prob}) 
in the limit of $n \rightarrow \infty$, and the same scaled angular momenta $\ell/n$ and $\ell'/n$.

By using Euler-Maclaurin formula, the summation of Equation (\ref{prob}) is approximated by an 
integration such that
\begin{equation}
\label{int_lim}
\lim_{n\rightarrow \infty} P_{\ell'\ell}^{(n)} = 2\ell' n
\int_0^1
\left\{\begin{array}{ccc}
\ell' & \ell & L\\
j & j & j
\end{array} \right\}^2 \; H_{jL}^2(\chi ) \; d\left( \frac{L^2}{n^2}\right)  +
{\cal O}\left(\frac 1n\right)
\end{equation}
where $H_{jL}$ is the generalized character associated with the irreducible 
representation of the rotation group \citep{star2}, defined by:
\begin{equation}
H_{j L} (\chi ) = L!\;
\sqrt{\frac{(2 j + 1)(2 j -L)!}{(2 j + L +1)!}}
\left( 2 \sin \chi \right)^L  C^{(1+L)}_{2 j - L}\left(\cos \chi \right)\;,
\end{equation}
in terms of ultraspherical polynomials $C^{(\alpha)}_n$. In this section
$j = (n+1)/2$.

The classical limit of the 6-j symbol was first discussed by \citet{wign}, 
based on the physical interpretation that the square of the 6-j symbol gives the 
quantum probability of coupling 3 angular momenta $j_1, j_2$ and $j_3$ to the 
total angular momentum $j$. The corresponding classical probability is then linked 
to the volume of a tetrahedron that has $j_1$, $j_2$ and $j_3$ as edges joining at 
a vertex, and $j$, $j_{12}$ and $j_{23}$ as the other edges, opposing that vertex. 
Here $j_{12}$ and $j_{23}$ are the intermediate coupling angular momenta. 
This provides only an estimate, and the agreement is quantitative only on average over
neighboring angular momenta. Heuristic arguments led \citet{ponz} to an improved 
formula which takes into account the oscillations inside the classically 
allowed region, but which fails in the vicinity of the turning points.
A rigorous derivation of these results and an uniform approximation valid
over a large range of values of the angular momentum, including classically
forbidden ones, is given by \citet{schu}.

The semiclassical approximation for the square of the 6-j symbol is
\begin{equation}
\left\{\begin{array}{ccc}
\ell' & \ell & L\\
j & j & j
\end{array} \right\}^2 \approx
\left[\frac 1{\sqrt{12\pi V}}\cos (\Theta + \pi/4)\right]^2 =
\frac 1{24\pi V} (1 + \sin 2 \Theta)
\end{equation}
and zero when $V^2 < 0$ \citep{ponz}.
The semiclassical phase $\Theta$ is defined by $\Theta = \sum_{i< k} j_{ik} \theta_{ik}$ 
where $j_{ik}$ is the length of the edge opposed to edge $ik$ and $\theta_{ik}$ is the
dihedral angle between faces intersecting on the edge opposed to $ik$. 
Wigner's estimate is obtained when the oscillatory factor, from the above formula, is neglected.

Using notation $z=L/n$, $\cos \eta_1 = \ell/n$ and $\cos \eta_2 = \ell'/n$,
the volume of the tetrahedron is calculated as
\begin{equation}
\label{6JSC}
V = \frac 13 j^3\left\{ \left[ \sin^2 (\eta_1 + \eta_2) -z\right]
\left[z - \sin^2 (\eta_1 - \eta_2)\right]\right\}^{1/2}
\end{equation}
The generalized character function $H_{jl}$ is a solution of a differential equation, 
\begin{equation}
\label{H_ode}
\frac{d^2 H_{jL}(\omega)}{d\omega^2} + 2 \cot \omega \frac{d H_{jL}(\omega)}{d\omega}+
\left[ 4j(4j+1) - \frac{L(L+1)}{\sin^2 \omega}\right] H_{jL}(\omega) = 0\;,
\end{equation}
derived from the corresponding differential equation for Gegenbauer polynomials.
A more convenient form for Equation (\ref{H_ode}) is obtained by setting 
$f = \sin \omega\; H_{jL}(\omega )$ and $\omega = \pi/2 +x$.
The resulting equation
\begin{equation}
f'' (x) + \left[ (2 j + 1)^2 -\frac{L(L+1)}{\cos^2 x}\right] f(x) = 0
\end{equation}
is a one-dimensional Schr\"odinger equation for a particle moving between
$-\pi/2 < x < \pi/2$ in a symmetric potential. A simple WKB solution for
this equation can be constructed to obtain the approximation
\begin{equation}
\label{HSC}
H_{jL}^2(\chi ) \approx \frac 1{2 \sin\chi}
\left[ \sin^2 \chi - (L/n)^2 \right]^{-1/2}\;.
\end{equation}

Finally, by using the approximation Equation (\ref{6JSC}) for the 6-j symbol and
Equation (\ref{HSC}) for the generalized character in summation Equation (\ref{int_lim}),
a semiclassical approximation for the angular momentum changing probability
is obtained as
\begin{equation}
P^{SC}(\ell/n, \ell'/n, \chi ) =
\frac{\ell'}{\pi n^2 \sin\chi} \int
dz\;\left[\left(z - \sin^2(\eta_1 - \eta_2)\right)
\left( \sin^2(\eta_1 + \eta_2) - z\right)\left( \sin^2\chi - z \right)
\right]^{-1/2} \;.
\end{equation}
This is precisely the same formula obtained from purely classical phase space
arguments in \citet{star1}. The integration is limited to a proper domain, where
the argument of the square root function is positive, otherwise, in the classical 
limit, the probability for transition is zero.
Moreover, the integral can be calculated in terms of the complete
elliptic integral
$K(m) = \int_0^{\pi/2} (1- m \sin^2 x)^{-1/2}\;dx$ 
(formula 3.131.4 of \citet{grad}), to finally obtain the semiclassical approximation 
Equation (\ref{probSC}).



\clearpage

\begin{thebibliography}{}
%

\bibitem[Abrines(1966)]{abrines} Abrines, R., \& Percival, I. A. 1966, Proc. Phys. Soc., 88, 861 

\bibitem[Beigman(1995)]{beig} Beigman, I.L., \& Lebedev, V.S. 1995, Phys. Rep., 250, 95

\bibitem[Bergemann(2010)]{berg} Bergemann, M. 2010, in Uncertainties in Atomic Data and How They Propagate 
in Chemical Abundances, ed. V. Luridiana, J. Garc\'ias Rojas, \& A. Manchado (Instituto de Astrofisica de Canarias), 
in press (arXiv:1104.1640)

\bibitem[Brocklehurst(1970)]{brok} Brocklehurst, M. 1970, MNRAS, 148, 417 

\bibitem[Cavagnero(1995)]{cava} Cavagnero, M.J. 1995, Phys. Rev. A, 52, 2865 

\bibitem[Chluba et al.(2007)]{chlu1} Chluba, J., Rubi\~no-Martin, J.A., \& Sunyaev, R.A. 2007, MNRAS, 374, 1310 

\bibitem[Chluba et al.(2010)]{chlu2} Chluba, J., Vasil, G.M., \& Dursi, L.J. 2010, MNRAS, 407, 599 

\bibitem[Dalgarno(1983)]{dalgarno} Dalgarno, A, 1983, in Rydberg States of Atoms and Molecules, ed. 
R.F. Stebbings \&  F.B. Dunning (Cambridge University Press), 1

\bibitem[Demkov et al.(1974)]{demkov} Demkov, Yu.N., Ostrovskii, V.N., \& Solov'ev E.A. 1974, 
Zh. Eksp. Teor. Fiz. 66, 125 (Sov. Phys. JETP 39, 57)

\bibitem[Dutta et al.(2001)]{dutta} Dutta S.K., Feldbaum, D., Walz-Flannigan, A., Guest, J.R., 
\& Raithel G.2001, Phys. Rev. Lett., 86, 3993

\bibitem[Flannery \& Vrinceanu(2003)]{star3} Flannery, M.R. \& Vrinceanu, D. 2003, Int. J. Mass Spectr., 223, 473

\bibitem[Gabrielse(2005)]{gabrielse} Gabrielse, G. 2005, Adv. At. Mol. Opt. Phys., 50, 155 

\bibitem[Gradshteyn(2000)]{grad} Gradshteyn, I.S,. \&  Ryzhik, I.M. 2000, 
Table of Integrals, Series and Products (San Diego: Academic Press), 250

\bibitem[Hillier(2011)]{hill} Hillier, D.J. 2011, Ap\&SS, 336, 87

\bibitem[Luridiana et al.(2003)]{luri} Luridiana, V., Peimbert, A., Peimbert M., \& Cervi\~no, M. 
2003, ApJ, 592, 846

\bibitem[Mashonkina(2009)]{mash2} Mashonkina, L. 2009, Phys. Scr.,T 134, 014004 

\bibitem[Mashonkina(1996)]{mash1} Mashonkina, L.I., 1996, in ASP Conf. Ser. 108, 
Model Atmospheres and Spectrum Synthesis, ed. S.J. Adelman, F. Kupka, \& W.W. Weiss (San Francisco, CA:ASP), 140 

\bibitem[Otsuka et al.(2011)]{otsu} Otsuka, M., Meixner M., Riebel D., Hyung S., Tajitsu A., \&  Izumiura H. 2011, 
ApJ, 729, 39 

\bibitem[Pengelly \& Seaton(1964)]{peng} Pengelly, R.M., \&  Seaton, M.J. 1964, MNRAS, 127, 165

\bibitem[Percival \& Richards(1977)]{percival} Percival, I.C., \& Richards, D. 1977, J. Phys. B 10, 1497

\bibitem[Pipher \& Terzian(1969)]{piph} Pipher, J.L., \& Terzian, Y. 1969, ApJ, 155, 475 

\bibitem[Pohl et al.(2008)]{pohl} Pohl, T., Vrinceanu, D., \& Sadeghpour, H.R.  2008, Phys. Rev. Lett., 100, 223201

\bibitem[Ponzano \& Regge(1968)]{ponz} Ponzano, G.,\& Regge, T. 1968, in Semiclassical Limit of Racah Coefficients,
Spectroscopic and Group Theoretical Methods in Physics (Amsterdam, North-Holland), 100

\bibitem[Przybilla \& Butler(2004)]{przy} Przybilla, N., \& Butler, K. 2004, ApJ., 609, 1181 
 
\bibitem[Sampson(1977)]{samp} Sampson, D.H. 1977, J. Phys. B, 10, 749 

\bibitem[Samuelson(1970)]{samu} Samuelson, R.E. 1970, J. Atmos. Sci., 27, 711 

\bibitem[Schlag \& Levine(1997)]{schlag} Schlag, E.W., \& Levine, R.D. 1997, Comments At. Mol. Phys., 33, 159

\bibitem[Schulten \& Gordon(1975)]{schu} Schulten, K., \& Gordon, R.G. 1975, J. Math. Phys., 16, 1971 

\bibitem[Smith \& Chupka(1995)]{smit} Smith, J.M., \& Chupka, W.A. 1995, J. Chem. Phys., 103, 3436 

\bibitem[Sun \& MacAdams(1993)]{sun} Sun, X., \& MacAdams, K.B. 1993, Phys. Rev. A, 47, 3913 

\bibitem[Vrinceanu \& Flannery(2000)]{star} Vrinceanu, D., \& Flannery, M.R. 2000, Phys. Rev. Lett., 85, 4880

\bibitem[Vrinceanu \& Flannery(2001a)]{star1} Vrinceanu, D., \& Flannery, M.R. 2001, Phys. Rev. A, 63, 032701

\bibitem[Vrinceanu \& Flannery(2001b)]{star2} Vrinceanu, D., \& Flannery, M.R. 2001, J. Phys. B, 34, L1 

\bibitem[Wigner(1959)]{wign} Wigner, E. P., 1959, Group Theory (New York: Academic Press), 100

\end{thebibliography}
\end{document}